\documentclass[12pt]{article}
\def\be{\begin{equation}}
\def\ee{\end{equation}}
\def\bea{\begin{eqnarray}}
\def\eea{\end{eqnarray}}

\def\e{\,{\rm e}}

\def\L{{\cal L}}
\def\mphi{m_{\phi}}

\def\chip{\chi_{+}}
\def\chim{\chi_{-}}

\def\half{\frac{1}{2}}
\def\quart{\frac{1}{4}}
\def\lam4{\frac{\lambda}{4}}

\def\nonum{\nonumber}

\def\lsim{{\
\lower-1.2pt\vbox{\hbox{\rlap{$<$}\lower5pt\vbox{\hbox{$\sim$}}}}\ }} 
\def\gsim{{\
\lower-1.2pt\vbox{\hbox{\rlap{$>$}\lower5pt\vbox{\hbox{$\sim$}}}}\ }}

\begin{document}
\baselineskip 0.65cm
\begin{titlepage}

\begin{flushright}
    ICRR-Report-360-96-11\\
    hep-ph/9603317\\
    March 1996
\end{flushright}

\vspace{0.5cm}

\begin{center}
    {\Large\bf Restriction to Parametric Resonant Decay}\\
    {\Large\bf after Inflation}\\
    \vspace{1.5cm} 
    {\large S.~Kasuya$^{*}$ and
    M.~Kawasaki$^{\dagger}$}\\
    \vspace{1cm}
    {\it Institute for Cosmic Ray Research, University of Tokyo,
    Tanashi, Tokyo 188, Japan\\
    $^{*}$e-mail:kasuya@icrr.u-tokyo.ac.jp\\
    $^{\dagger}$e-mail:kawasaki@icrr.u-tokyo.ac.jp}
\end{center}

\vspace{3.5cm}

\begin{abstract}
    We study parametric resonant decay of inflaton field with emphasis
    on its physical meaning. We show that the parametric resonance is
    indeed an induced process, which means that the more numbers of
    produced particles, the more inflaton field decays. We also
    consider the dissipative effects of produced particles and find
    that the dissipation reduces the resonant decay rate of inflaton
    field.
\end{abstract}

\end{titlepage}
\newpage

\section{Introduction}

Inflation \cite{guth,sato} is the most successful theory to solve many
problems of the standard ``hot big bang'' Universe. In the
inflationary Universe scenario the universe is expanded exponentially
by the vacuum energy of some scalar field $\phi$ (inflaton).  For this
scenario to work, the radiation-dominated Universe must be recovered
after inflationary stage. Therefore, some process is necessary to
transfer the vacuum energy to relativistic particles. This process is
called reheating. 

The old version of reheating theory was first considered
in~\cite{abbott, dolgov} for the new inflationary theory \cite{new},
and it can be applied to the chaotic inflation \cite{chaotic}. After
inflation, inflaton field oscillates near the bottom of its effective
potential. Inflaton field decays into other (lighter) particles due to
its coupling to others. The reheating temperature can be estimated as
$T_{RH}\simeq10^{-1}\sqrt{\Gamma_{tot}M_p}$~\cite{linde}, based on the
single-particle decay. However, recent investigation revealed that the
drastic decay of inflaton field $\phi$ occurs in the first stage of
reheating \cite{kofman, shtanov, boyanovsky, yoshimura, kaiser,
traschen}. This 
stage is called preheating~\cite{kofman}. At first, inflaton field
$\phi$ explosively decays into some bosons $\chi$, whose spectrum is
far from equilibrium. After $\chi$-particles collide with each others
or decay into lighter particles, thermal equilibrium can be achieved 
and reheating is complete.

The equation for the particles produced can be described by Mathieu
type equation, which has instability solution in some regions of
parameters (instability bands). If the relevant parameters stay in the
instability bands long enough, the solution will explosively grow,
which means that the number of particles created becomes exponentially
large so that the parametric resonant decay takes place very
efficiently~\cite{kofman, shtanov, boyanovsky, yoshimura, kaiser}.

 From the different viewpoint, the physical meaning of parametric
resonant decay could be considered as an ``induced'' decay, which is
similar to an induced emission of photons in the two energy level
system of atoms, in the sense that the presence of produced particles
stimulates inflaton field to decay into those particles. Thus this
phenomenon is peculiar to those particles that obey Bose-Einstein
statistics, bosons, the number of which in each mode $k$ can increase
exponentially. This is the reason that parametric resonant decay into
fermions cannot occur due to Pauli's exclusion
principle~\cite{dol-kiri, kofman, shtanov, boyanovsky}.

If the parametric resonance is an induced process, the parametric
resonant decay of inflaton into bosons are likely to be suppressed by
the processes of scattering or decay of the produced
particles~\cite{dol-kiri, shtanov}. The reason is that dissipation
(the particle scattering, decay and annihilation) will tend to
decrease the mean occupation numbers $N_k$ in each particular mode so
that the induced decay will be reduced, and, as a result, these
numbers cannot grow so large during the resonance period.

Therefore, in the present paper we study the parametric resonant decay
in the narrow resonance region~\footnote{%
We study onlyin the narrow resonance region since we expect that the
suppresion will be unimportant in the broad resonance region.}
with emphasis on its physical meaning
and consider the effect of dissipation.  In section 2, we will make
clear that the parametric resonant decay is an induced effect, and we
will see the restriction to that decay due to effects of dissipation
in section 3. Section 4 contains our conclusion and discussion.

\section{Parametric Resonant Decay as an Induced Effect}

In this paper we assume the interaction between inflaton field $\phi$
and $\chi$-particles as $\L_{int}=-\sigma\phi\chi^2$ and the effective
potential of inflaton field as $V(\phi)=\frac{1}{2}\mphi^2\phi^2$. The
similar results can be obtained for other interactions and also for 
$\frac{\lambda}{4}\phi^4$ potential.  The mode equation for $\chi$
field becomes
\be
   \ddot{\chi}_k+[\omega_k^2+2\sigma\phi_0\sin(\mphi t)]\chi_k=0,
   \label{ind-eom}
\ee
where $\omega_k^2=k^2+m_{\chi}^2$ with $m_{\chi}$ being the mass of
the $\chi$-particle, $\phi_0$ is the amplitude of the inflaton field,
and the expansion of the Universe is neglected.

We can expand $\chi$ field as
\be
   \chi=\frac{1}{(2\pi)^{3/2}}\int d^3k
   [ a(k)\chim(t) \e^{i\vec{k}\vec{x}}
   +a^{\dagger}(k) \chip(t) \e^{-i\vec{k}\vec{x}} ],
   \label{chi-exp}
\ee
where $a^{\dagger}(k)$ and $a(k)$ are creation and annihilation
operators, respectively, with commutation relation:
$[a(k),a^{\dagger}(k^{\prime})]=\delta(\vec{k}-\vec{k^{\prime}})$.
$\chi_{\pm}$ are the positive and negative frequency solutions of
Eq.(\ref{ind-eom}), which are equal to the positive and negative
frequency asymptotic solutions ($=\exp(\pm k_0 t)/\sqrt{2k_0}$) if
$\chi$ was free field.

Using Eq.~(\ref{chi-exp}) the energy density of the $\chi$-particle can be
calculated  as
\bea
   \langle 0 \,|\, \rho_{\chi} \,|\, 0 \rangle & = & \langle 0 \,|\, 
   \half\dot{\chi}^2+\half\Omega_k^2\chi^2 \,|\, 0 \rangle \nonum \\ 
   & = & \frac{1}{(2\pi)^3}\int d^3k 
   \left[ \half|\dot{\chi}_-|^2+\half\Omega_k^2|\chim|^2 \right],
\eea
where we take the vacuum expectation value.  Thus the energy density
for the $\chi$-particle with momentum $k$ is given by 
\be
   \rho_{\chi}^k=\half|\dot{\chi}_-|^2+\half\Omega_k^2|\chim|^2. 
\ee
Then the time evolution of $\rho_{\chi}^k$ is described as
\be 
    \dot{\rho}^k_{\chi} = \frac{d}{dt} \left(
        \half|\dot{\chi}_-|^2+\half\Omega_k^2|\chim|^2 \right)
    =\half\frac{d(\Omega_k^2)}{dt}|\chi_-|^2,
    \label{rho-k}
\ee
where the Eq.~(\ref{ind-eom}) is used in the second step.

Hereafter, we assume that $\sigma \phi_0 \ll \mphi^2$ which
corresponds to so-called narrow resonance region. This assumption
allows us to take perturbative analysis.~\footnote{%
Analyzing in the narrow resonance region, we can regard that the
amplitude of inflaton field is approximately constant.} 
First, we will show that our analysis is general so that it does not
depent on the exponential growth nature of the solution to
(\ref{ind-eom}). If we write the solution as
\be
    \chim=A\cos(\omega_k t),
\ee
where $A$ is a slowly time dependent, complex function, the energy
density can be given as follows:
\bea
    \rho_k & = & \half|\dot{\chi}_k|^2+\half|\chim|^2,  \nonum \\
           & = & \half|A|^2\omega_k^2\cos^2(\omega_k t)
               + \half|A|^2\omega_k^2\sin^2(\omega_k t), \nonum \\  
           & = & \half|A|^2\omega_k^2.
           \label{gen-rho-k}
\eea
 From the equation (\ref{rho-k}), we get
\bea
    \dot{\rho}_k & = & \half 2\sigma\phi_0\mphi\cos(\mphi t)
                      |A|^2 \cos^2(\omega_k t), \nonum \\
                 & = & \half \sigma\phi_0\mphi|A|^2 \cos(\mphi t)
                             [1+\cos(2\omega_k t)],
\eea
and taking the average over one period of oscillations of inflaton
field $\phi$, the rhs of this equation will not vanish when the
resonance condition $\omega_k=\mphi/2$ holds so that, 
\bea
     \dot{\rho}_k & \simeq & \quart \sigma\phi_0\mphi|A|^2, \nonum \\ 
                  & = & \frac{2\sigma\phi_0}{\mphi}\rho_k,
    \label{gen-rho-dot}
\eea
where (\ref{gen-rho-k}) is used in the second line. 

$\dot{\rho}^k_{\chi}$ can be identified as the production rate of the
$\chi$-particle with momentum $k$ due to parametric resonant decay of
inflaton field $\phi$.  Since the production rate is proportional to
the energy density of produced $\chi$-particles ( hence its occupation
numbers in each $k$ mode), we can see that this is an induced effect.
That is, if many particles are created in the particular phase space,
further production will be ``induced''.

Now we will find and use the explicit form of the solution for
(\ref{ind-eom}) to obtain the previous result. Hamiltonian is a
diagonalized operator for the free field, but in the presence of
interaction, it is not diagonalized in terms of $a$ and
$a^{\dagger}$. It can be diagonalized at any instant of time by means
of Bogolyubov transformations, which give the transition from the free 
field creation and annihilation operator, $a^{\dagger}$ and $a$, to
time dependent operators, $b^{\dagger}(t)$ and $b(t)$ (which preserve
the commutation relation),
\bea
   b(t) & = & \alpha(t) a + \beta^*(t) a^{\dagger}, \nonum \\
   b^{\dagger}(t)& = & \beta(t) a + \alpha^*(t) a^{\dagger},   
   \label{qft-bogolyubov}
\eea
and $\alpha(t)$ and $\beta(t)$ can be written as~\cite{shtanov}
\bea
    \alpha & = & \frac{\e^{i\int\Omega_k dt}}{\sqrt{2\Omega_k}}
    (\Omega_k\chim+i\dot{\chi}_{-}), \nonum \\ 
    \beta & = & \frac{\e^{-i\int\Omega_k dt}}{\sqrt{2\Omega_k}}
    (\Omega_k\chim-i\dot{\chi}_{-}), 
    \label{alpha-beta}
\eea
where $\Omega_k^2=\omega_k^2+2\sigma\phi$, and the initial conditions
for $\alpha$, $\beta$ are
\be
   |\alpha(0)|=1, \hspace{8mm} \beta(0)=0.
   \label{alpha-beta-init}
\ee

Then the form of the solution can be written as, to the lowest
order~\cite{shtanov},  
\be
    \chim=\frac{1}{\sqrt{2\omega_k}}
    [X\cos(\omega_k t)-Y\sin(\omega_k t)],
\ee
where $X$, $Y$ are complex weakly time-dependent functions. From
Eqs.(\ref{alpha-beta}) and (\ref{alpha-beta-init}) the initial
conditions for $\chi$ become
\be 
    i\dot{\chi}_{-}(0)=\omega_k \chim(0),
    \hspace{5mm} |\chim(0)|=\frac{1}{\sqrt{2\omega_k}},
    \label{chim-init}
\ee
which lead to the initial values for $X$ and $Y$:
\be
    X(0)=\e^{i\delta}, \hspace{5mm} Y(0)=i\e^{i\delta},
\ee
where $\delta$ is some phase which is not important and is set 
to zero below. 

We are only interested in the case where the resonance condition holds:
\be
    \omega_k \approx \frac{\mphi}{2}.
    \label{res-cond}
\ee
Therefore, the approximate solutions for $X$ and $Y$ are given by
\bea
    X & \simeq & \frac{1+i}{2}\e^{\mu t}+\frac{1-i}{2}\e^{-\mu t}, 
    \nonum \\ 
    Y & \simeq & \frac{1+i}{2}\e^{\mu t}-\frac{1-i}{2}\e^{-\mu t},
    \label{x-y}
\eea
where $\mu=\mphi^{-1}\sqrt{\sigma^2\phi_0^2-\Delta^2}$, and
$\Delta=\omega_k^2-\mphi^2/4$. Therefore we obtain the squares of the
amplitude of $\chim$ and $\dot{\chi}_{-}$ in the form
\bea
   |\chim|^2 & = &\frac{1}{2\omega_k} 
   [\cosh(2\mu t)-\sinh(2\mu t)\sin(2\omega_k t)],
   \label{chi-2}\\
   |\dot{\chi}_{-}|^2 & = & \frac{1}{2\omega_k} \omega_k^2
   [\cosh(2\mu t)+\sinh(2\mu t)\sin(2\omega_k t)],
   \label{chi-dot-2}
\eea
Taking the average over one period of oscillations of inflaton field
$\phi$ and using Eqs.~(\ref{chi-2}) and (\ref{chi-dot-2}), we obtain
the averaged $\rho_{\chi}^k$ and $\dot{\rho}^k_{\chi}$:
\bea
    \langle \rho_{\chi}^k \rangle & \simeq &
    \frac{\omega_k}{2} \cosh(2\lambda t) 
    \simeq \frac{\mphi}{4} \cosh(2\lambda t) \\
    \langle \dot{\rho}^k_{\chi} \rangle
    & \simeq & \sigma\phi_0\sinh(2\mu t)
    \langle \sin(\mphi t)\sin(2\omega_k t)\rangle 
    = \half \sigma\phi_0\sinh(2\mu t),
\eea
where the non-vanishing value can be achieved only when the resonance
condition (\ref{res-cond}) holds.  Therefore, the following relation
is found to be satisfied:
\be
    \dot{\rho}^k_{\chi}
    =\frac{2\sigma\phi_0}{\mphi}\tanh(2\mu t)\rho^k_{\chi}
    \simeq \frac{2\sigma\phi_0}{\mphi}\rho^k_{\chi} 
    \hspace{8mm}({\rm as}\hspace{3mm} \mu t : {\rm large}).
    \label{rho-k-2}
\ee
Here we get the same result as (\ref{gen-rho-dot}). 

The total production rate is obtained by summing up all the mode:
\bea
    \dot{\rho}_{\chi} 
    & = & \int \frac{\omega_k^2}{2\pi^2} d\omega_k \rho^k_{\chi} 
    \nonum \\[0.6em] 
    & \simeq & \int \frac{\omega_k^2}{2\pi^2} d\omega_k 
    \half\sigma\phi_0\sinh(2\int^{\omega_k} \mu 
    \left|\frac{dt}{d\omega_k^{\prime}} \right| d\omega_k^{\prime}) 
    \nonum \\[0.6em]
    & \simeq & \frac{H}{2\pi^2} \left( \frac{\mphi}{2} \right)^4 
    \half \cosh \left( \frac{2\pi\sigma^2\phi_0^2}{H\mphi^3} \right),
\eea
where $\mu$ is regarded as adiabatically changing variable in the
resonance band and the integration is taken over only inside the
resonance band in the second line and, in the third, we use the
resonance condition ( $\omega_k \approx \mphi/2$ ) and $H$ is the
Hubble's constant. Using the relation
$\dot{\rho}=\Gamma_{\chi}^{(res)}\rho_{\phi}$ ( energy conservation ),
we can calculate the resonant decay rate given by
\bea
    \Gamma^{(res)}_{\chi} & \simeq & \frac{H}{2\pi^2 \rho_{\phi}} 
    \left( \frac{m_{\phi}}{2} \right) ^4 \half \cosh 
    \left( \frac{2\pi\sigma^2\phi_0^2}{Hm_{\phi}^3} \right) 
    \nonum \\[0.6em]
    & \simeq & \frac{Hm_{\phi}^2}{64\pi^2\phi_0^2}
    \exp \left( \frac{2\pi\sigma^2\phi_0^2}{Hm_{\phi}^3} \right) .
    \label{ind-gamma-res}
\eea
This is identical to the result of ref.~\cite{shtanov} 
(when $\int \mu dt$ is large).  

As mentioned before, we can obtain the similar results for other forms
of interaction between the inflaton and the $\chi$-particle or for
$\lam4 \phi^4$ potential.~\footnote{%
The behavior of inflaton field can be approximated with a sinusoidal
function since its amplitude is small in the narrow resonance region.} 
Here, we only show the results corresponding to (\ref{rho-k-2}). For
the interaction $\L_{int}=-g\phi^2\chi^2$, we get 
\be
    \dot{\rho}^k_{\chi} \simeq
    \frac{g\phi_0^2}{2\mphi}\rho^k_{\chi}, 
\ee
and for the theory $V=\lam4\phi^4$, we obtain 
\bea
    \dot{\rho}^k_{\chi}
    & \simeq & \frac{2\sigma}{c\sqrt{\lambda}}\rho^k_{\chi},
    \hspace{7mm} {\rm for} \hspace{3mm} \L_{int}=-\sigma\phi\chi^2, \\
    & \simeq & \frac{g\phi_0}{2c\sqrt{\lambda}}\rho^k_{\chi},      
    \hspace{5mm} {\rm for} \hspace{3mm} \L_{int}=-g\phi^2\chi^2,
\eea
where $c \sim {\cal O}(1)$.

\section{Restriction due to dissipative effects}

As shown in the previous section, the parametric resonant decay of
inflaton field is an induced effect of produced bosonic particles. If
there are large occupation numbers of produced bosons, further creation
proportional to their occupation numbers will occur.  The crucial point
is that a lot of bosons can occupy in the same phase space.

On the other hand, if particles in particular phase space decay into
other particles or collide with each other, occupation numbers of
bosons in that particular phase space cannot grow very fast. As a 
result, the resonant decay will be much suppressed. In other words,
the resonance band becomes  narrower in the presence of decay or
scattering processes.

We begin with the mode equation for the $\chi$ field which is produced by
the decay of inflaton field $\phi$. One convenient way to take into
account the decay of the field $\chi$ is to introduce a dumping term
in the mode equation.~\footnote{%
To include the dissipative process in the mode equation is not so
simple as given in this paper. The precise treatment of dissipation is
very complicated subject and  beyond the scope of the present
paper. However, we believe that the simple argument given here is
reasonable in the physical point of view.}
Then the equation becomes a type of equation for
damping oscillator with an external oscillating force for the theory
$V=\half\mphi^2\phi^2$ and $\L_{int}=-\sigma\phi\chi^2$ (we can obtain
the similar results for different types of effective potential $V(\phi)$
or interaction Lagrangian $\L_{int}$),  
\be
    \ddot{\chi}_k + \Gamma_d \dot{\chi}_k +
    [\omega_k^2+2\sigma\phi_0\sin(m_{\phi}t)]\chi_k=0,
    \label{decay-eom-1}
\ee
where $\Gamma_d$ is the decay rate for the $\chi$ field.
Then we obtain the solution $\chi_k$ for Eq.(\ref{decay-eom-1}) 
in the following form:
\be
    |\chim|^2=\frac{\e^{-\Gamma_d t}}{2\omega_k}
    [\cosh(2\mu t)-\sinh(2\mu t)\sin(2\omega_k t)].
\ee
The evolution of the energy density of the $\chi$-particle in $k$ mode
becomes
\bea
   \dot{\rho}^k_{\chi} & = & \frac{d}{dt} \left( 
       \half|\dot{\chi}_-|^2+\half\Omega_k^2|\chim|^2 \right)
   \nonum \\[0.6em]
   & = & \half\frac{d(\Omega_k^2)}{dt}|\chi_-|^2 - \Gamma_d |\chim|^2
   \nonum \\[0.6em]
   & \simeq & \frac{2\sigma\phi_0}{\mphi}\rho^k_{\chi} 
   - \Gamma_d \rho^k_{\chi},  
   \label{rho-k-diss}
\eea
where we use Eq.(\ref{decay-eom-1}) in the second line and, in the third, 
take the average over one period of oscillations of inflaton field
$\phi$. The first term is the usual resonant production from decay of
inflaton field $\phi$ and the second is the energy decrease due to the 
decay of the $\chi$-particle. Therefore, the energy density of
the $\chi$-particle in the mode $k$ is reduced by the factor
$\e^{-\Gamma_d t}$ compared with the case of no dissipation, and hence the
considerable decrease of the resonant decay rate of inflaton field. 

Summing over mode $k$, we obtain the total resonant production of energy
density of the $\chi$-particle:
\bea
   \dot{\rho}_{\chi} 
   & = & \int \frac{\omega_k^2}{2\pi^2} d\omega_k 
   \half\frac{d(\Omega_k^2)}{dt}|\chi_-|^2  \nonum \\[0.6em]
   & \simeq & \int \frac{\omega_k^2}{2\pi^2} d\omega_k 
   \quart\sigma\phi_0\exp(\int^{\omega_k} (2\mu-\Gamma_d) \left| 
       \frac{dt}{d\omega_k^{\prime}} \right| d\omega_k^{\prime}) 
   \nonum \\[0.6em]
   & \simeq & \frac{H}{2\pi^2} \left( \frac{\mphi}{2} \right)^4 
   \quart\exp \left[ \frac{4\sigma^2\phi_0^2}{H\mphi^3} 
       \left( \sin^{-1}
           \sqrt{1-\left( \frac{\mphi\Gamma_d}{2\sigma\phi_0} \right)^2}
           -\frac{\mphi\Gamma_d}{2\sigma\phi_0}
           \sqrt{1-\left( \frac{\mphi\Gamma_d}{2\sigma\phi_0} \right)^2} 
       \right) \right], \nonum \\
   & & 
   \label{rho-dot}
\eea
where, in the third line, we take into account that the resonance band
width gets narrower due to dissipation:  
$\Delta=\sigma\phi_0 \rightarrow \sigma\phi_0 
\sqrt{1-\left( \frac{\mphi\Gamma_d}{2\sigma\phi_0} \right)^2}$. If
$\Gamma_d < \sigma\phi_0/\mphi (\simeq \mu)$, the production
rate becomes approximately,  
\be
    \dot{\rho}_{\chi} \simeq \frac{H}{2\pi^2} 
    \left( \frac{\mphi}{2} \right)^4 
    \quart\exp \left[ \frac{2\pi\sigma^2\phi_0^2}{H\mphi^3} 
        \left( 1-\frac{2\mphi\Gamma_d}{\pi\sigma\phi_0} \right) \right].
\ee
Now we can see that the resonant production of $\chi$-particles is
reduced due to dissipative effects, and hence the reduction of resonant
decay rate of inflaton field $\phi$. 

Simpler explanation can be given as follows. The naive integration 
in the exponent in Eq.(\ref{rho-dot}) becomes
\bea
    \int (2\mu-\Gamma_d) dt 
    & = &  \int 2\mu dt - \int \Gamma_d dt \nonum \\
    & \simeq &  \int 2\mu dt - \Gamma_d \tau_{res}, 
    \label{naive-int}
\eea
where we regard $\Gamma_d$ as constant, and $\tau_{res}$ is the
duration of the parametric resonance (the time while staying in the
resonance band). If $\tau_{res}>\Gamma_d^{-1}$, i.e., the lifetime of
the $\chi$-particle is shorter than the resonant duration, the
particle does not stay in the particular phase space long enough that
the occupation number in that particular mode cannot grow so fast,
which results in less resonant production.

\section{Conclusion}

We investigate the parametric resonant decay of inflaton field $\phi$
and the dissipative effects of particles produced after the chaotic
inflation. The equation for the particles produced can be described by
Mathieu type equation, which has instability solution in some regions
of parameters (instability bands). If the frequencies stay in the
instability bands long enough, the solution will explosively grow, so
that the number of created particles becomes exponentially large.
Therefore the previous authors~\cite{kofman, shtanov, boyanovsky,
yoshimura, kaiser} regarded that the parametric resonant decay is
important.

In this article, we considered this phenomena from another point of
view.  The resonant decay is an induced effect of bosonic particles.
If there are large occupation numbers of produced bosons, further
creation proportional to their numbers will occur.  A lot of bosons
can occupy in the same phase space while one fermion can occupy in the
particular phase space. This is the reason that the inflaton field
does not decay explosively into fermions (Pauli's exclusion principle)
\cite{dol-kiri, kofman, shtanov, boyanovsky}.

Then, in the case of decaying into bosons, if the dissipative effects
prevent the occupation numbers of bosons from growing very fast, the
resonant decay will be much suppressed.  
The dissipative processes involve collisions between particles created, 
and their decay or annihilation into other lighter particles.  

The simplest way to look for this effect is compare the duration of
resonance (the time that resonance frequency stays in the instability
band) with the lifetime of produced particles. If the resonant
duration is shorter than the particle lifetime, resonance occurs
considerably. In the opposite case, parametric resonance should be
extremely suppressed. We believe that this criterion is general.

The dissipative effects are taken into account phenomenologically by
$\Gamma_d\dot{\chi}_k$ term in the equation for the mode $k$ of the
field $\chi$, although it is difficult to obtain the term
$\Gamma_d\dot{\chi}_k$ from the first principle. 
Besides, decay rates of particles depend on its energy which is
redshifted by the expansion of the Universe, so that the investigation
is much more complicated. However, we expect that the result will
remain the same qualitatively.  

There are some cosmological implication from this effects. First,
those particles that are created by the decay of inflaton field should 
weakly interact with themselves or with other particles. Otherwise,
the dissipative processes become effective so as to reduce the
resonant decay of inflaton field considerably. Second, the reheating
temperature will not be so high because thermalization takes for a
long time.  
 
Finally we must mention in closing that our simple analytical
investigation in this article for both parametric resonance and
dissipative effects  will be limited to only in the narrow
resonance region where the effects of back-reaction to the
amplitude of inflaton filed is rather small. There are some papers
studying broad resonance, which is rather
complicated~\cite{kofman,boyanovsky,yoshimura,linde95}. However, we
expect that the basic physical process for broad resonance is the same
as narrow one. Therefore, an induced nature of parametric resonant
decay will also hold qualitatively in the broad resonace region so
that the dissipation tends to suppress the resonant decay in the same
way. (It is model-dependent whether or not this suppresion becomes
important, and we expect that it will be less efficient in the broad
resonance region, since the resonance duration will be much shorter
than the lifetime of created particles.) 

\vspace{0.5cm}
\noindent
The authors are grateful to T.~Yanagida for useful discussions.

\end{document}